\documentclass[english,aps,preprint]{revtex4}
\usepackage[T1]{fontenc}
\usepackage[latin9]{inputenc}
\usepackage{array}
\usepackage{amstext}
\usepackage{graphicx}

\makeatletter

\DeclareFontEncoding{LGR}{}{}
\DeclareTextSymbol{\~}{LGR}{126}
\providecommand{\tabularnewline}{\\}

\@ifundefined{textcolor}{}
{%
 \definecolor{BLACK}{gray}{0}
 \definecolor{WHITE}{gray}{1}
 \definecolor{RED}{rgb}{1,0,0}
 \definecolor{GREEN}{rgb}{0,1,0}
 \definecolor{BLUE}{rgb}{0,0,1}
 \definecolor{CYAN}{cmyk}{1,0,0,0}
 \definecolor{MAGENTA}{cmyk}{0,1,0,0}
 \definecolor{YELLOW}{cmyk}{0,0,1,0}
 }

\makeatother

\makeatother

\usepackage{babel}

\begin{document}

\title{Resonant production of the fourth family slepton at the LHC}

\author{O. \c{C}ak\i{}r}

\email{ocakir@science.ankara.edu.tr}

\author{S. Kuday}

\email{kuday@science.ankara.edu.tr}

\affiliation{Department of Physics, Ankara University, Faculty of Sciences, Ankara,
Turkey}

\author{\.{I}.T. \c{C}ak\i{}r}

\email{tcakir@mail.cern.ch}

\affiliation{Department of Physics, CERN, Geneva, Switzerland}

\author{S. Sultansoy}

\email{ssultansoy@etu.edu.tr}

\affiliation{Physics Division, TOBB University of Economics and Technology, Ankara,
Turkey }

\affiliation{Institute of Physics, Academy of Sciences, Baku, Azerbaijan}

\begin{abstract}
The resonant production of the fourth family slepton $\tilde{l}_{4}$
via R-parity violating interactions of supersymmetry at the Large
Hadron Collider has been investigated. We study the decay mode of
$\tilde{l}{}_{4}$ into the fourth family neutrino $\nu_{4}$ and
$W$ boson. The signal will be a like-sign dimuon and dijet if the
fourth family neutrino has Majorana nature. We discuss the constraints
on the R-parity violating couplings $\lambda$ and $\lambda'$ of
the fourth family charged slepton at the LHC with the center of mass
energies of 7, 10 and 14 TeV. 
\end{abstract}

\maketitle

Existence of the fourth family follows from the basics of the standard
model (SM) and actual mass spectrum of third family fermions (see
\cite{Sahin11} and references therein). 
Recent studies \cite{Maltoni00,He01,Kribs07,Bobrowski09,Chanowitz09,Alok10,Chanowitz10,Erler10,Eberhardt10,Cobanoglu10}
on the allowed parameter space for the fourth family fermions from
precision electroweak data show that this space is large enough. The
experimental limits on the masses of the fourth family quarks from
Collider Detector at Fermilab (CDF) are: $m_{u_{4}}>335$ GeV at 95\%
CL. \cite{Conway}, $m_{d_{4}}>338$ GeV at 95\% CL. \cite{Aaltonen}.
There are also limits on the masses of the fourth family leptons \cite{Nakamura10}:
$m_{l_{4}}>100$ GeV, $m_{\nu_{4}}>90$ ($80$) GeV for Dirac (Majorana)
neutrinos. On the other hand, the partial wave unitarity leads to
an upper bound 700 GeV for fourth SM family fermion masses \cite{Chanowitz78}.

Obviously, if the fourth SM family exists in nature then the minimal
supersymmetric standart model (MSSM) should be enlarged to include
fourth family superpartners. The inclusion of the fourth SM family
into MSSM is straightforward \cite{Carena96} (we denote minimal supersymmetric
standart model with three and four families as MSSM3 and MSSM4, respectively).
A search for supersymmetry (SUSY) is significant part of the physics
program of TeV scale colliders. As mentioned in \cite{Ari11}, it
is difficult to differentiate MSSM3 and MSSM4 at hadron colliders,
because the light superpartners of the third and fourth family quarks
has almost the same decay chains if the $R$-parity is conserved.
For this reason the pair production of fourth family charged sleptons
at future $e^{+}e^{-}$ colliders has been proposed in \cite{Ari11}
to differentiate the MSSM with three and four families.

The R-parity is defined as $R=(-1)^{3(B-L)-2S}$, where $B,\, L$
and $S$ are the baryon number, lepton number and spin, respectively.
It is a useful assignment for the phenomenology, because all the SM
particles and Higgs boson have even R-parity, while all the sfermions,
gauginos and Higgsinos of the supersymmetry have odd R-parity. The rich phenomenology 
of the MSSM becomes even richer if R-parity is violated (see \cite{Barbier05} and references therein). 
Concerning MSSM4, R-parity violation (RPV) could provide opportunity to 
differentiate MSSM3 and MSSM4 at hadron colliders.

The R-parity violating part of the MSSM superpotential is given by

\begin{equation}
W_{RPV}=\lambda_{ijk}L_{i}L_{j}E_{k}^{c}+\lambda'_{ijk}L_{i}Q_{j}D_{k}^{c}+\lambda''_{ijk}U_{i}^{c}D_{j}^{c}D_{k}^{c}\label{eq:1}\end{equation}
where $L(E)$ is an $SU(2)$ doublet (singlet) lepton superfield and
$Q(U,D)$ is (are) an $SU(2)$ doublet (singlet) quark superfield(s),
and indices $i,j,k$ denote flavour. The coefficients $\lambda_{ijk}$
and $\lambda''_{ijk}$ corresponds to the lepton number violating
and baryon number violating couplings, respectively. The second term
in Eq. \ref{eq:1} allows the resonance production of sleptons at
hadron colliders.  

In this work, we consider the resonant production of fourth family
slepton via the process $pp\rightarrow\tilde{l_{4}}^{+}X\rightarrow\nu_{4}\mu^{+}X$
and followed by the decay $\nu_{4}\rightarrow W^{-}\mu^{+}$ for the
Majorana nature of neutrino at the Large Hadron Collider (LHC) with
$\sqrt{s}=7$, 10 and $14$ TeV. Assuming that $W$-boson decays hadronically,
the signal will be seen in detector as $\mu^{+}\mu^{+}jj$. 

The RPV supersymmetric trilinear interaction terms for the charged
fourth family slepton can be written as

\begin{equation}
L{}_{RPV}=\lambda_{i4k}\tilde{l}_{4L}\bar{l}_{kR}\nu{}_{i}+\lambda_{ij4}\tilde{l}_{4R}^{*}\bar{\nu}_{i}^{c}l_{jL}-\lambda_{4jk}\tilde{l}_{4L}\bar{l}_{kR}\nu{}_{j}-\lambda_{ij4}\tilde{l}_{4R}^{*}\bar{\nu}_{j}^{c}l_{iL}-\lambda'_{4jk}\tilde{l}_{4L}\bar{q}_{kR}q_{jL}+h.c.\label{eq:2}
\end{equation}
where $\tilde{l}_{4L(R)}$ is the fourth family slepton field, $q_{L(R)}$
is the left-handed (right-handed) quark field, and indices $i,j,k$
denote flavour.

The mass matrix of the fourth family charged sleptons in the $(\tilde{l}_{4L},\tilde{l}_{4R})$
basis is given by
\begin{equation}
M_{\tilde{l}_{4}}^{2}=\left(\begin{array}{cc}
m_{\tilde{l}_{4L}}^{2} & a_{l_{4}}m_{l_{4}}\\
a_{l_{4}}m_{l_{4}} & m_{\tilde{l}_{4R}}^{2}\end{array}\right)\label{eq:3}
\end{equation}
where $m_{\tilde{l}_{4L}}^{2}=M_{\tilde{L}_{4}}^{2}+m_{l_{4}}^{2}-m_{Z}^{2}$$\cos2\beta(\frac{1}{2}-\sin^{2}\theta_{W})$;
$m_{\tilde{l}_{4R}}^{2}=M_{\tilde{E}{}_{4}}^{2}+m_{l_{4}}^{2}-m_{Z}^{2}$$\cos2\beta\sin^{2}\theta_{W}$;
$a_{l_{4}}=A_{l_{4}}-\mu\tan\beta$, and $A_{l_{4}}$ is the Higgs-fourth
family charged lepton trilinear parameter (the notation of \cite{Bartl97}
is used).

The mass eigenstates $\tilde{l}_{4l}$ and $\tilde{l}_{4h}$ are related
to $\tilde{l}_{4L}$ and $\tilde{l}_{4R}$ by

\begin{equation}
\left(\begin{array}{c}
\tilde{l}{}_{4l}\\
\tilde{l}_{4h}\end{array}\right)=\left(\begin{array}{cc}
\cos\theta_{\tilde{l_{4}}} & \sin\theta_{\tilde{l_{4}}}\\
-\sin\theta_{\tilde{l_{4}}} & \cos\theta_{\tilde{l_{4}}}\end{array}\right)\left(\begin{array}{c}
\tilde{l}_{4L}\\
\tilde{l}_{4R}\end{array}\right)\label{eq:4}
\end{equation}
with the eigenvalues

\begin{equation}
m_{\tilde{l}_{4(l,h)}}^{2}=\frac{1}{2}(m_{\tilde{l}{}_{4L}}^{2}+m_{\tilde{l}{}_{4R}}^{2})\mp\frac{1}{2}\sqrt{(m_{\tilde{l}{}_{4L}}^{2}-m_{\tilde{l}{}_{4R}}^{2})^{2}+4a_{l_{4}}^{2}m_{l_{4}}^{2}}\label{eq:5}
\end{equation}
and the mixing angle $\theta_{\tilde{l}_{4}}$ is given by

\begin{equation}
\cos\theta_{\tilde{l}_{4}}=\frac{-a_{l_{4}}m_{l_{4}}}{\sqrt{(m_{\tilde{l}{}_{4L}}^{2}-m_{\tilde{l}{}_{4l}}^{2})^{2}+a_{l_{4}}^{2}m_{l_{4}}^{2}}}\label{eq:6}
\end{equation}
 As seen from Eq. (\ref{eq:5}), $\tilde{l}{}_{4l}$ is expected to
be the lightest charged slepton because of large value of $m_{l_{4}}$.

\begin{figure}
\includegraphics{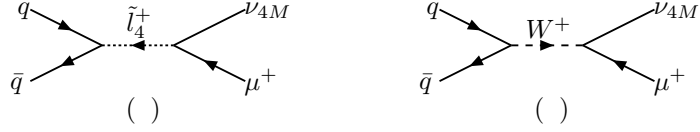}
\caption{Feynman diagrams of subprocess $q\bar{q'}\rightarrow\nu_{4}\mu^{+}$:
(a) Signal (b) Background\label{fig:1}}
\end{figure}

The hadronic cross section for the process $pp\to\tilde{l}_{4l}^{+}X\to\mu^{+}\bar{\nu}X$
is defined by

\begin{eqnarray}
\sigma=\sum_{i,j}\int\int dx_{1}dx_{2}\hat{\sigma}_{part}(x_{1}x_{2}s)[f_{i}(x_{1},Q^{2})f_{j}(x_{2},Q^{2})+f_{i}(x_{2},Q^{2})f_{j}(x_{1},Q^{2})]\label{eq:7}
\end{eqnarray}
where $x_{1}$ and $x_{2}$ are the fractions of parton momentum to
proton momentum for two proton beams, $s$ is the square of the center
of mass energy, and $Q$ is the factorization scale. For the subprocess
shown in Fig.\ref{fig:1}(a), the partonic cross section $\hat{\sigma}_{part}(\hat{s})$
is calculated as

\begin{equation}
\hat{\sigma}_{part}(\hat{s})=\sum_{jk}\frac{C_{F}(\lambda'{}_{4jk}^{eff}\lambda_{442}^{eff})^{2}(\hat{s}-m_{\nu_{4}}^{2})^{2}}{16\pi\hat{s}[(\hat{s}-m_{\tilde{l}_{4}}^{2})^{2}+m_{\tilde{l}_{4}}^{2}\Gamma_{\tilde{l_{4}}}^{2})]}\label{eq:8}
\end{equation}
where $m_{\tilde{l}_{4}}$and $m_{\nu_{4}}$ are the masses of fourth
family charged slepton and fourth family neutrino, respectively; $C_{F}$
is the color factor, and the effective couplings are defined as $\lambda^{eff}(\lambda'^{eff})=\cos\theta_{\tilde{l}_{4}}\lambda(\lambda')$.

\begin{figure}
\includegraphics[scale=0.6]{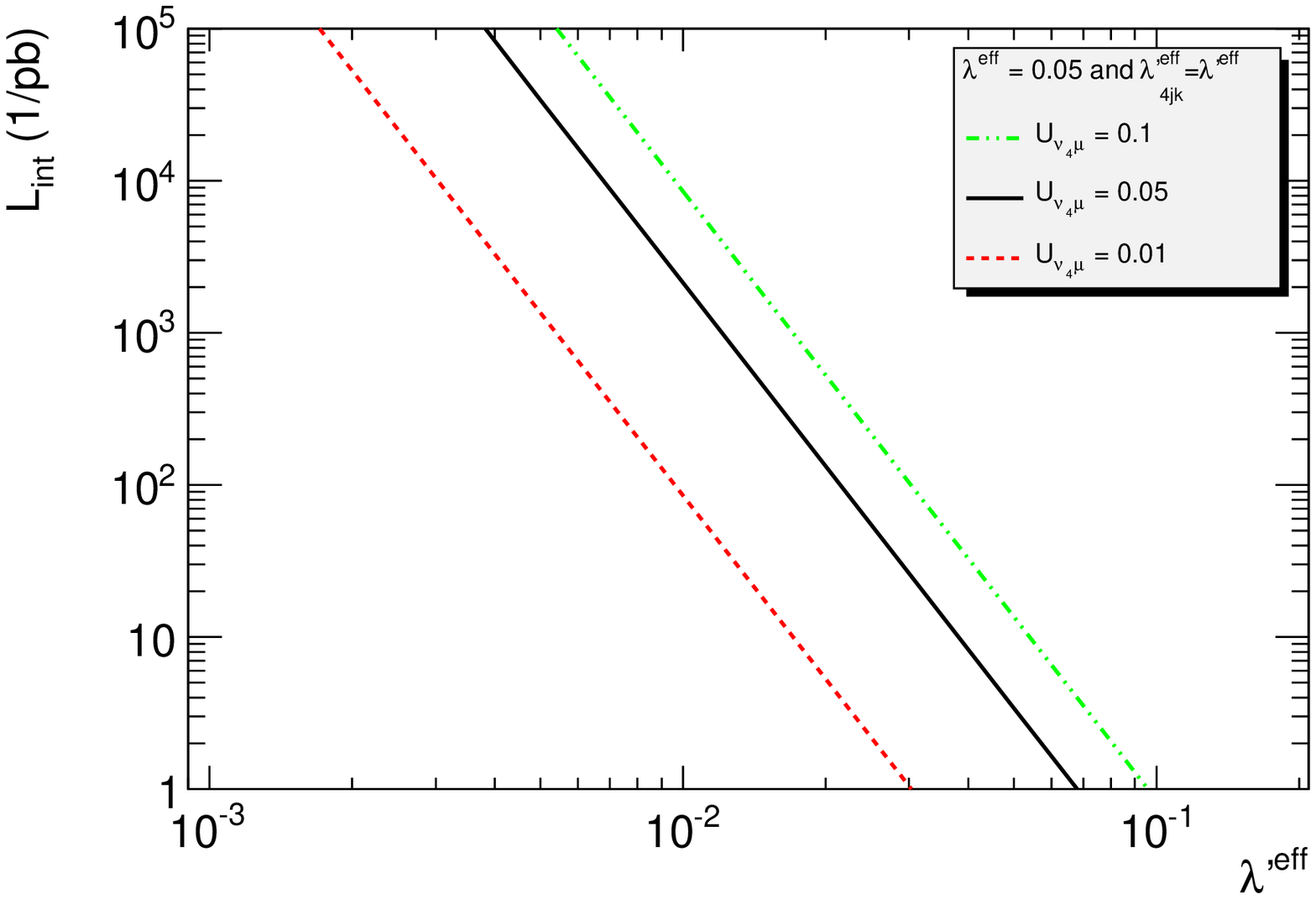}
\caption{Integrated luminosity versus $\lambda^{'eff}$ for 3$\sigma$
significance at $\sqrt{s}=7$ TeV.\label{fig:2} }
\end{figure}

For numerical calculations we implement the vertices from interaction
Lagrangian (Eq. 2) into CompHEP \cite{CompHEP} with the CTEQ6M \cite{CTEQ02}
parton distribution functions. Masses of the fourth family (Majorano)
neutrino and charged lepton are taken as $m_{\nu_{4}}=100$ GeV and
$m_{\tilde{l}_{4}}=300$ GeV, respectively. It should be noted that
the main background for our signal, namely $\mu^{+}\mu^{+}jj$, will
come from the fourth SM family itself (see Fig. \ref{fig:2}(b)).
This background is proportinal to $|U_{\nu_{4}\mu}|^{2}$ . Recent
analysis of PMNS matrix elements in the presence of a fourth generation
showed that $|U_{\nu_{4}\mu}|<0.115$\cite{Lacker10}.

The signal cross sections at LHC with $\sqrt{s}=7,\,10$ and $14$
TeV are given in Tables \ref{tab:1}, \ref{tab:2} and \ref{tab:3},
respectively. In numerical calculations, we assume $\lambda_{442}^{eff}$=0.05,
in columns 2-7 corresponding $\lambda_{4jk}^{'eff}$
is equal to 0.05 and remaining ones are zero, in the last columns
all $\lambda_{4jk}^{'eff}$ s are equal to 0.05. Using
$|U_{\nu_{4}\mu}|=0.05$ and $m_{\nu_{4}}=100$ GeV, we also calculate
the background cross sections as $0.016$, $0.024$ and $0.035$ pb
for LHC with $\sqrt{s}=7,\,10$ and $14$ TeV, respectively. We use
the branching ratio $BR(\nu_{4}\to\mu^{+}W^{-})=0.34$ which is predicted
within the parametrization \cite{Ciftci05} compatible with the experimental
data on the masses and mixings in the leptonic sector. In the last
two rows of Table \ref{tab:1}, \ref{tab:2} and \ref{tab:3}, we
present the statistical significance for the signal observations and
required integrated luminosity for reaching $3\sigma$.

\begin{table}
\caption{Cross sections and significance depending on effective RPV couplings
at $\sqrt{s}=7$ TeV with $L_{int}=1$ $fb^{-1}$\label{tab:1}}

\begin{tabular}{|>{\centering}m{1.2in}|c|c|c|c|c|c|>{\centering}m{2cm}|}
\hline 
\noalign{\vskip\doublerulesep}
 & $\lambda_{411}^{'eff}$  & $\lambda_{412}^{'eff}$  & $\lambda_{421}^{'eff}$  & $\lambda_{413}^{'eff}$  & $\lambda_{422}^{'eff}$  & $\lambda_{423}^{'eff}$  & $\lambda_{4jk}^{'eff}$ \tabularnewline
\hline
\hline 
\noalign{\vskip\doublerulesep}
$\sigma_{S}(pb)$  & 0.15  & 0.11  & 0.02  & 0.06  & 9.3x10$^{-3}$  & 3.9x10$^{-3}$ & 0.35\tabularnewline
\hline 
\noalign{\vskip\doublerulesep}
$S/\sqrt{B}$  & 22  & 16 & 3 & 8.8 & 1.4 & 0.6 & 52\tabularnewline
\hline 
\noalign{\vskip\doublerulesep}
$L_{int}(pb^{-1})$ for 3$\sigma$  & 18.8  & 35 & 1x10$^{3}$ & 120 & 5x10$^{3}$  & 2.8x10$^{4}$  & 3.4\tabularnewline
\hline
\end{tabular}
\end{table}

\begin{table}
\caption{The same as for Table 2 but for $\sqrt{s}=10$ TeV and $L_{int}=1$00
$fb^{-1}$ \label{tab:2}}
\begin{tabular}{|c|c|c|c|c|c|c|>{\centering}p{2cm}|}
\hline 
 & $\lambda_{411}^{'eff}$  & $\lambda_{412}^{'eff}$  & $\lambda_{421}^{'eff}$  & $\lambda_{413}^{'eff}$  & $\lambda_{422}^{'eff}$  & $\lambda_{423}^{'eff}$  & $\lambda_{4jk}^{'eff}$ \tabularnewline
\hline
\hline 
$\sigma_{S}(pb)$  & 0.24  & 0.19  & 0.033  & 0.11  & 0.021  & 9.6x10$^{-3}$ & 0.6\tabularnewline
\hline 
$S/\sqrt{B}$  & 287 & 225 & 38.7 & 130 & 25.2 & 11.4 & 716\tabularnewline
\hline 
$L_{int}(pb^{-1})$ for 3$\sigma$  & 11 & 17.8 & 6x10$^{2}$ & 53.4 & 1.4x10$^{3}$ & 6.9x10$^{3}$ & 1.753\tabularnewline
\hline
\end{tabular}
\end{table}

\begin{table}
\caption{The same as for Table 2 but for $\sqrt{s}=14$ TeV \label{tab:3}}
\begin{tabular}{|c|c|c|c|c|c|c|>{\centering}p{2cm}|}
\hline 
 & $\lambda_{411}^{'eff}$  & $\lambda_{412}^{'eff}$  & $\lambda_{421}^{'eff}$  & $\lambda_{413}^{'eff}$  & $\lambda_{422}^{'eff}$  & $\lambda_{423}^{'eff}$  & $\lambda_{4jk}^{'eff}$ \tabularnewline
\hline
\hline 
$\sigma_{S}(pb)$  & 0.36  & 0.29  & 0.061  & 0.177  & 0.042  & 0.02 & 0.96\tabularnewline
\hline 
$S/\sqrt{B}$  & 350 & 290 & 60 & 175 & 41.4 & 19.7 & 946\tabularnewline
\hline 
$L_{int}(pb^{-1})$ for 3$\sigma$  & 7.23 & 10.7 & 247 & 30 & 525 & 2.3x10$^{3}$  & 1\tabularnewline
\hline
\end{tabular}
\end{table}

\begin{figure}
\includegraphics[scale=0.6]{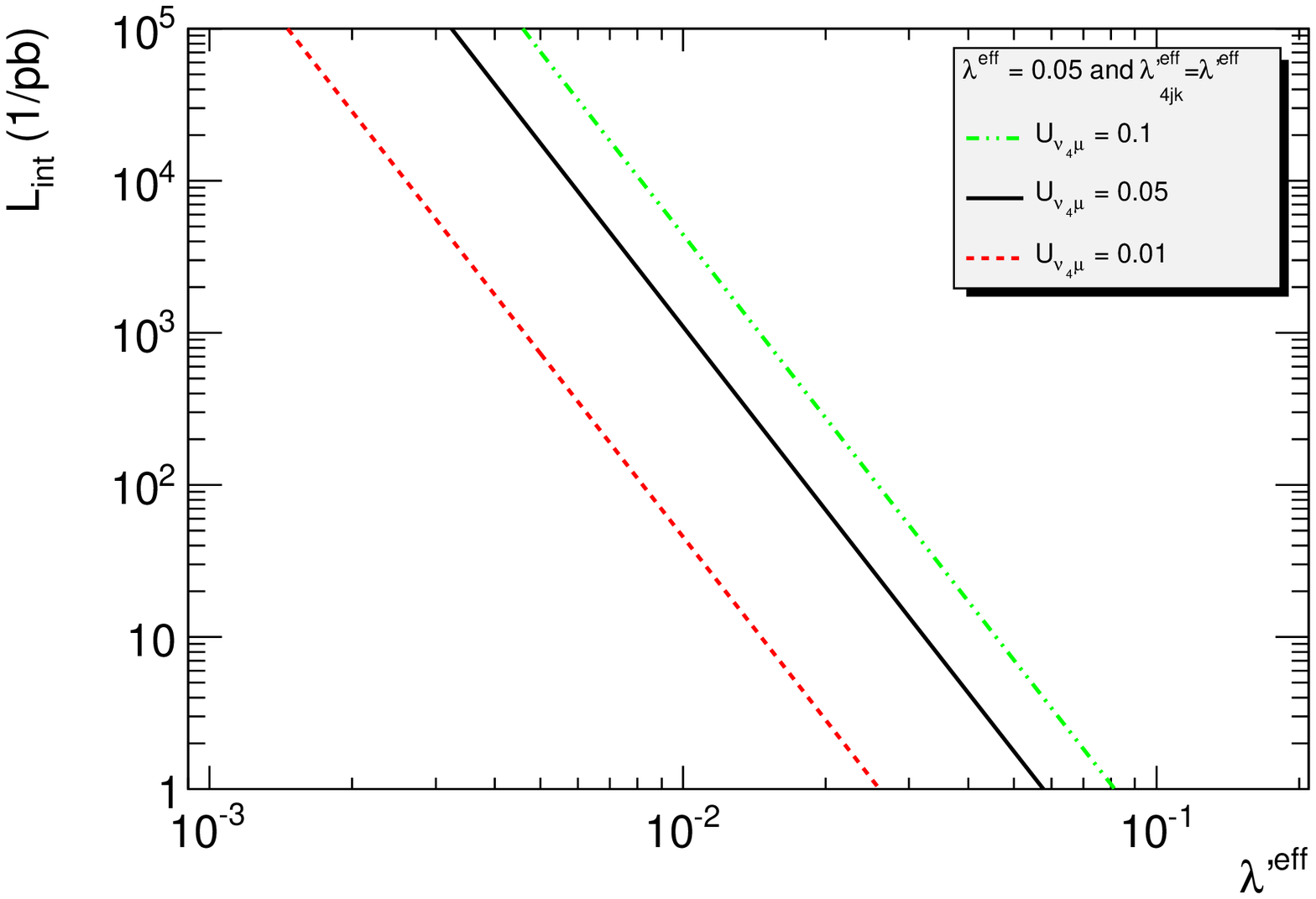}
\caption{Integrated luminosity versus $\lambda^{'eff}$ for 3$\sigma$
significance at $\sqrt{s}=10$ TeV.\label{fig:3} }
\end{figure}

\begin{figure}
\includegraphics[scale=0.6]{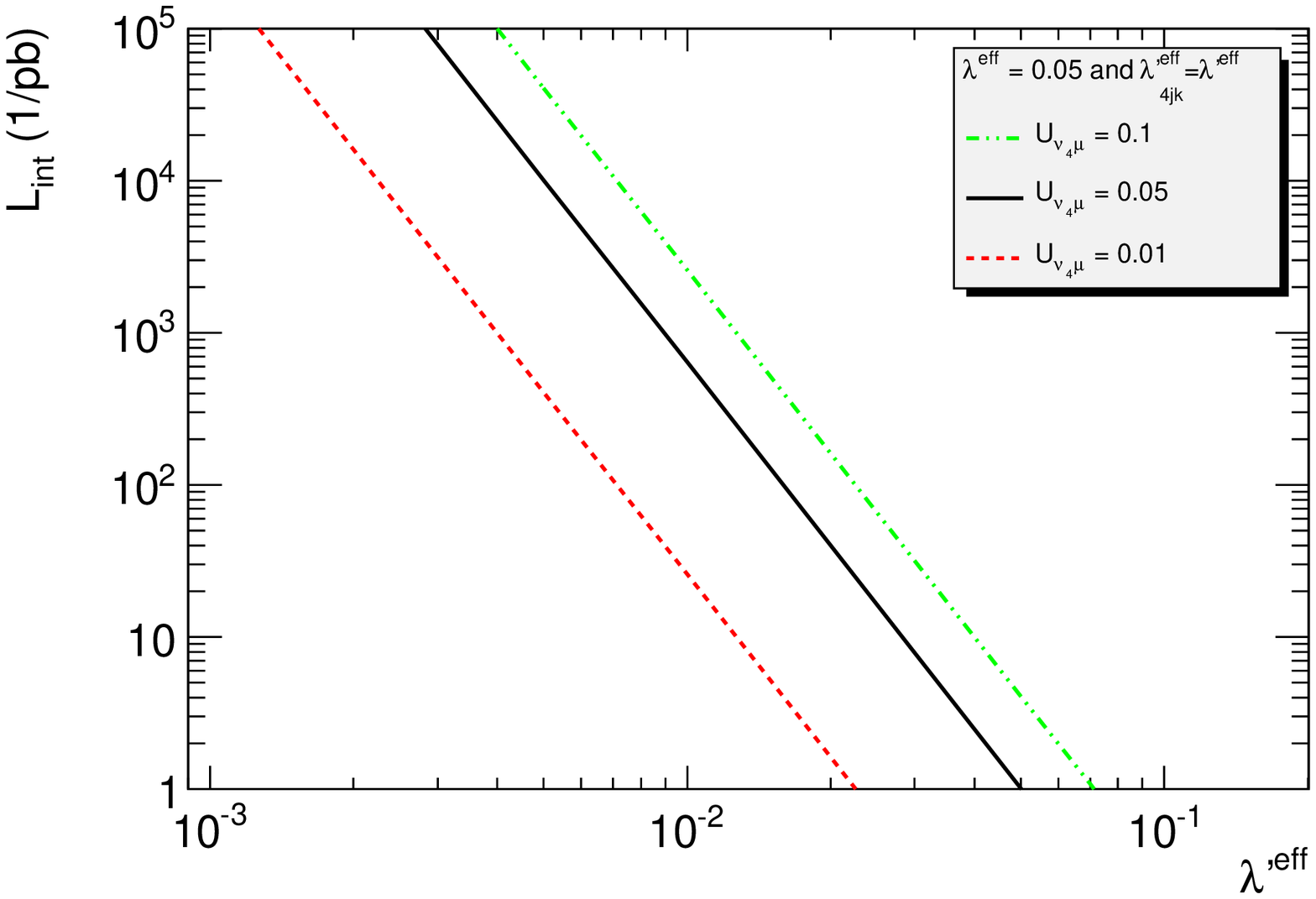}
\caption{Integrated luminosity versus $\lambda^{'eff}$ for 3$\sigma$
significance at $\sqrt{s}=14$ TeV.\label{fig:4} }
\end{figure}

In figures \ref{fig:2}, \ref{fig:3} and \ref{fig:4} we plot integrated
luminosity needed for 3$\sigma$ significance reach depending
on $\lambda_{4jk}^{'eff}\equiv\lambda^{'eff}$ for three different values
of $|U_{\nu_{4}\mu}|$ assuming $m_{\nu_{4}}=100$ GeV, $m_{\tilde{l}_{4}}=300$
GeV and $\lambda_{442}^{eff}=0.05$. In Table \ref{tab:4} we present
observable values of $ $$ $$\lambda^{'eff}$ for 3$\sigma$
observation at the LHC runs with $\sqrt{s}=7,\,10$ and $14$ TeV.

\begin{table}
\caption{Achievable value of $\lambda^{'eff}$ for 3$\sigma$ observation.
\label{tab:4}}
\begin{tabular}{|c|c|c|c|}
\hline 
$|U_{\nu_{4}\mu}|$$ $ & $\sqrt{s}=7$ TeV, $L_{int}=1\, fb^{-1}$ & $\sqrt{s}=10$ TeV, $L_{int}=100\, fb^{-1}$ & $\sqrt{s}=14$ TeV, $L_{int}=100\, fb^{-1}$\tabularnewline
\hline
\hline 
0.1 & 0.017 & 0.0048 & 0.0040\tabularnewline
\hline 
0.05 & 0.010 & 0.0032 & 0.0028\tabularnewline
\hline 
0.01 & 0.0045 & 0.0015 & 0.0012\tabularnewline
\hline
\end{tabular}
\end{table}

The analysis shows that fourth family 
sleptons can be measured with 3$\sigma$ significance having $10^{-3}$ for $\lambda^{'eff}$ 
and $10^{-2}$ for $|U_{\nu_4 \mu}|$ at the LHC (14 TeV) with 100 fb$^{-1}$. 

In conclusion, we have studied the resonance production of fourth family sleptons 
through R-parity violating couplings at the LHC energies, and it could be the 
first manifestation of the MSSM4 at the LHC.

\begin{acknowledgments}
This work is partially supported by Turkish Atomic Energy Authority (TAEK) and 
State Planning Organization (DPT).
\end{acknowledgments}

\end{document}